\documentstyle[aaspp4,12pt]{article}
\def\etal{{\it et al.}}
\def\eg{{\it e.g.}}
\def\ie{{\it i.e.}}
\def\cf{{\it cf.}}
\def\deg{{^\circ}}
\def\Mpc{{\mbox{Mpc}}}
\def\hMpc{{h^{-1}\mbox{Mpc}}}
\begin{document}
\title{Two Dimensional Topology of Large Scale Structure in the Las Campanas
Redshift Survey}

\author{ Wesley N. Colley \footnote{Supported by the Fannie and John Hertz
Foundation, Livermore, CA 94551-5032}}
\affil{Princeton University Observatory, Peyton Hall, Princeton, NJ 08544}

\begin{abstract}
We have measured the topology (genus) of the density distribution of
large-scale structure observed in the Las Campanas Redshift Survey (LCRS).  The
LCRS is complete to magnitude 17.5, and contains nearly 24000 galaxies with
median redshift of 30000 km/s.  The large volume and large number of galaxies
allows sampling of nearly 100 independent structures with which to compute the
genus topology, a vast improvement over previous studies.  We find that the
genus is consistent with a random-phase Gaussian distribution of initial
density fluctuations, as would be produced naturally in inflationary models.
When we combine these results with the genus measurements of the COBE microwave
background fluctuations, we find that two orthogonal projections of the
three-dimensional distribution of initial density fluctuations are consistent
with Gaussian random-phase behavior, in agreement with standard inflationary
models.  Particular attention is given to statistical significance of the genus
test.
\end{abstract}


\section{Introduction}

The nature of the large-scale structure of the Universe is one of the
pre-eminent questions of modern astronomy.  Several prodigious observational
efforts have been made to survey this structure in the optical, (Geller \&
Huchra 1989) infrared (Moore \etal\ 1992), and the microwave (Smoot \etal\
1994).  The most recent of these efforts has been the Las Campanas Redshift
Survey (Shectman \etal\ 1996).  This survey includes roughly 24000 galactic
redshifts, which is a significant improvement over previous optical efforts
(\cf\ Geller and Huchra 1989).  The survey visibly includes many walls, voids
and filaments in each slice, which allows, perhaps for the first time, a
statistical understanding of the geometry of the large-scale structure.

While the surveyors themselves have measured in detail the power-spectrum seen
in the LCRS (Lin \etal\ 1996), we focus on the phase information in the sample.
Inflation predicts that the large-scale structure on physical scales larger
than the non-linear regime should derive from a Gaussian random-phase
distribution of quantum noise (Bardeen, Steinhardt \& Turner 1983).  In other
words, the fluctuations should have Gaussian distributed amplitudes at random
phase.

To test this hypothesis, we have employed the genus topology method of Gott
\etal\ (1987).  Melott \etal\ (1989) showed that the number of structures
(two-dimensional genus) above and below a threshold in a Gaussian random-phase
field has a simple analytic form.  We therefore measure the genus of the LCRS
and compare to this analytic form to assess the similarity of the large-scale
density distribution to a Gaussian random-phase field.

\section{The Las Campanas Redshift Survey}

The Las Campanas Redshift Survey (Shectman \etal\ 1996) is the most complete
optical redshift survey to date.  It contains 23697 redshifts, with a median
redshift of 30000 km/s.  The survey is contained in six slices of constant
declination, each $1.5\deg \times 80\deg$.  Three of the slices are at low
declination ($\delta = -3\deg, -6\deg, -12\deg$) in the north galactic cap
region; the other three are at high declination ($\delta = -39\deg, -42\deg,
-45\deg$) in the south galactic cap region (Shectman \etal\ 1996).  Plainly
visible in the slices are large walls and voids, similar to those seen in
previous surveys (\cf\ Geller and Huchra 1989).

The enormity of this effort makes more significant our topology study, since
the genus statistic requires many clustered galaxies to count as a single
overdense structure.  The vast number of galaxies in the LCRS allows
statistically significant results since the number of structures counted is so
large.

Redshift surveys provide an ideal source of data for topology statistics.
While the Cosmic Microwave Background provides a slice at constant radius at
large redshift ($z \sim 1000$), a redshift survey provides a slice at a
constant declination (polar angle).  The topology of the CMB has been computed
by Colley \etal\ 1996, who found the fluctuations therein to be consistent with
a Gaussian random-phase distribution.  The LCRS provides information in the
orthogonal direction, which makes more compelling the notion that the evidence
in two dimensions for Gaussian random-phase fluctuations extends to three
dimensions.  Studies of the 3-D topology in a variety of samples (\eg\ the CfA
Survey, Abell Clusters, and IRAS samples) have all been consistent with
Gaussian random phase initial conditions (Gott \etal\ 1989, Vogeley \etal\
1994, Gott, Rhoads and Postman 1994, Moore \etal\ 1992).  Of course, the Sloan
Digital Sky Survey will provide a redshift survey of comparable depth to the
LCRS, but over a full steradian of the sky (Lupton 1996).  As data from the
SDSS trickles in, we will begin to see a deep three-dimensional picture of
large-scale structure, and be able to assess the three-dimensional topology
readily.

\section{Assessing the Genus}

Slices in declination are not planar surfaces.  They are cones, which have
extrinsic, but no intrinsic curvature.  A cone can be cut radially and laid
flat on a plane with no distortion; it will look like a pizza with a slice
missing.  The size of the slice depends on the angle at the cone's apex.  If we
think of the cone's apex as the Earth, and the cone's edge as lying on the
celestial sphere on a circle of constant declination, we can compute the size
of the missing slice, by computing the arc distance on the sphere travelled
while tracing the edge of the cone.  The line-integral $\int\cos\delta d\alpha$
is simply $2\pi\cos\delta$, since declination, $\delta$, is constant.  Thus,
all that is mathematically required to map the cone onto the plane without
distortion is to use substitute $\alpha\cos\delta$ for $\alpha$ as the
azimuthal angle (where $\alpha$ is the right ascension).  After such a mapping
the three the high-declination slices, though they cover a wider range of right
ascension, travel approximately the same arc length on the celestial sphere as
do the lower declination slices, approximately $80\deg$.

Once we have correctly mapped the slices, we are able to produce map images on
which pixels corresponded to properly scaled physical units.  We first
construct the selection function which allows us to properly scale the
flux-limited ($m_{limit} \la 17.75$) sample so that it reflects galaxy density,
independent of flux, at a given radius.  First we find the maximum distance in
redshift space at which the $ith$ galaxy could be seen
\begin{equation}
D_{max,i} = 10^{0.2(m_{limit}-m_i)}\cdot cz/H_0
\end{equation}
The expected volume-density of galaxies as a function of radius is then
\begin{equation}
\rho_s(r) = {3\over{\Omega_s}} \sum_{D_{max,i}>r}{D_{max,i}^{-3}}
\end{equation}
(Gott \etal\ 1989) where $\Omega_s$ is the solid angle of the slice.  Since we
are interested in the two-dimensional structure, we require the expected
surface density of galaxies, $\sigma_s(r)$, which must grow linearly with
radius to account for the wedge-shaped profile of the slice.
\begin{equation}
\sigma_s(r) = Sr\rho_s(r)
\end{equation}
(Park \etal\ 1992) where $S$ is the ``shape factor,'' $(\sin\delta_h -
\sin\delta_l)/\cos\bar\delta$, where $\delta_u$ is the upper limit of
declination in the slice, $\delta_l$ is the lower limit, and $\bar\delta$ is
defined such that $\sin\bar\delta = (\sin\delta_u + \sin\delta_l)/2$
(so that $\bar\delta$ is the declination at which half the volume of the slice
lies to the north, and half to the south).

We cut an $80^\circ$ slice of this azimuthally symmetric selection function,
and smooth it.  We may then smooth the data, and divide it by the smoothed
selection function, which produces a map where the effects of the flux limit,
and edge effects have been minimized.  We truncate the slice at the radius
where the smoothing length, $\lambda$, equals $\sigma_s^{-1/2}$ to avoid shot
noise sampling effects.  To reduce these effects still further, we have
truncated the slices one smoothing length inside the cut-off radius.

We have chosen our smoothing length to ensure that the structures detected are
well within the linear regime.  In the linear regime, the topology of the
fluctuations is equal to the topology in the initial fluctuations because in
this regime fluctuations grow only in amplitude in place.  Since non-linear
effects typically become important on scales below about $8\Mpc$, we have
chosen a smoothing convolution kernel of $\exp(x^2 / 2\lambda^2)$, with
$\lambda = 20\hMpc$.  Furthermore, Matsubara (1996) has shown that edshift
peculiar velocities in the linear regime (\ie\ the Kaiser effect) do not
distort the topology of the fluctuations in redshift space.

Figure 1 provides a contour map of one of the smoothed, calibrated slices
($\delta = -39\deg$, width $1.5\deg$), with the galaxy locations overplotted.
The heavy lines are contours of high density; the lighter lines are contours of
low density; the dashed line is the median density contour.  We will discuss
the exact values of those contours in the next section.  For now we just wish
to illustrate that the approach we have described correctly identifies real
overdensities and voids in the data.  Also, many ``fingers of God'' are visible
in the data; these are indicators of non-linear effects on small scales (such
as virialized clusters). The structures visible in the contours, however, are
much larger than the fingers of God, which reassures us that we have smoothed
out most of the non-linear features in the data (the r.m.s. value of
$\delta\rho / \rho$, $\sigma_{20Mpc}$, = 0.462).

We also immediately note the large number of structures, critical to a
quantitative analysis via the genus statistic.  When compared with Park \etal\
(1992), which use the Geller \& Huchra (1989) survey, we find a vast
improvement in the number of structures detected (about a factor of 10), and
thus a much stronger lever-arm with which to perform topology statistics.


\section{The Genus of LCRS}

According to (Gott \etal\ 1990), we may define the genus $G$ of the excursion
set for a random density field on a plane as
\begin{equation}
\begin{array}{ll}
G = &\mbox{(number of isolated high-density regions)}- \\
~&   \mbox{(number of isolated low-density regions)}\\
\end{array}
\end{equation}
Equivalently, the genus can be defined as the total curvature of the contours.
Assuming a contour defines a differentiable curve $C$ on the map, its total
curvature is given by the integral
\begin{equation}
K = \int_C \kappa ds \equiv 2\pi G
\end{equation}
where $\kappa$ is the local curvature, $s$ parameterizes the curve, and $G$ is
the genus of a single contour.
An isolated overdense regions will contribute $+1$ to the total map genus and a
void (``hole'') in it will decrease the genus by 1.  Therefore the genus can be
considered as the total number of overdensities minus the total number of
voids.  In practice, contours may cross the edge of the survey region, in which
case the partial curves contribute non-integer rotation indices to the genus.
A two-dimensional random-phase Gaussian density field will generate a genus per
unit area
\begin{equation}
g \propto \nu e^{-\nu^2/2}
\end{equation}
where $\nu$ is the threshold value, above which a fraction, $f$ of the area
has a higher density
\begin{equation}
f = (2\pi)^{-1/2}
\int_\nu^\infty \exp(-x^2/2)dx
\end{equation}
(\cf\ also Adler 1981; Melott \etal\ 1989; Coles 1988; Gott \etal\ 1992; Park
\etal\ 1992).

In figure 1, we have plotted the contour map of the smoothed density
distributions in one of the slices, with contours at $\nu = \{-2,-1,0,1,2\}$.
On this map, many excursions can be seen at non-zero values of $\nu$, while the
median contour ($\nu = 0$) wanders through the map rather randomly.

The mean genus per unit area (averaged from all the estimates in each of the
six slices) is plotted as a function of $\nu$ in figure 2.  The best-fit
theoretical genus curve expected for a random-phase Gaussian distribution:
$g(\nu)\propto\nu\exp(-\nu^2/2)$ (eq. 3) is shown as a solid line.  The dotted
errorbars are the 68\% and 95\% confidence limits, estimated from the formal
Student's $t$-distribution for $n = 6$ (six slices) (Lupton 1993).

The fit of the genus curve to the data appears quite reasonable.  To be more
quantitative about this, we performed two tests, involving $\chi^2$-like
statistics.  First we computed $\tilde{\chi}^2$, which is different from
$\chi^2$ in that we have used the formal 1-$\sigma$ errors as calculated from
the six slices (recall these are $t$-distributed variates, not Gaussian),
\begin{equation}
\tilde{\chi}^2 = \sum_{i = 1}^{21}
{{(\bar{g}_i-\tilde{g}_i)^2}\over{\sigma_{\bar{g}_i,est}^2}} 
\end{equation}
Here, the sum runs over the 21 values of $\nu$ where we have measured the genus
(as shown in figure 2); $\bar{g}_i$ is the mean genus among the six slices at
each $\nu$-value; $\tilde{g}_i$ is the value of the fitted solid curve, and
$\sigma_{\bar{g}_i,est}$ is the standard error in the mean as estimated from
the six slices.  We ran $10^4$ simulations of 20 $t$-variates with $n = 6$ (one
less than 21, due to the one-parameter fit).  We then computed $\tilde{\chi}^2$
for these datasets, and found that our observed value of 46.6 fell at the 85\%
confidence interval (i.e. one would expect the value of $\tilde{\chi}^2$ to be
less than ours 85\% of the time, more 15\% of the time).  Thus, the results
were in agreement with the fitted curve $g(\nu) \propto \nu e^{\nu^2/2}$ within
the 95\% confidence interval.


One further concern was that the genus might be correlated slice-to-slice, or
among the different values of $\nu$.  Several of the slices are physically
proximate, since they come in two sets of three slices, whose minimum
separations are $3\deg$, which corresponds to our smoothing length of $20h^{-1}
\Mpc$ at a distance of $400\hMpc$.  One can, in fact, see that some of the
structures inside of $400\hMpc$ in neighboring slices are part of the same
three-dimensional structures.  To test the level at which this may corrupt our
genus error estimates, we computed the genus' covariance matrix elements
slice-to-slice.  We then summed these matrices over all the $\nu$ values to
calculate the total covariance in the genus between all slice pairs.  This
``total'' covariance matrix assumes the genus is a stationary process, which
may not be strictly true; however, given the apparent statistical similarity in
the errorbars at the various $\nu$-values, the assumption seems reasonable
enough to get a rough quantitative handle on the slice-to-slice covariance.  We
expect that if the physically proximate slices (the group of three in the
north, and the group of three in the south) contain substantial covariance, the
covariance matrix should look something like a two-by-two diagonal matrix,
rebinned into a six-by-six matrix.  The upper left three-by-three block would
contain the significant correlation of the three northern slices.  The
lower-right three-by-three block would contain the significant correlation of
the three southern slices.  The other two blocks, would contain no significant
correlation, since they come from slices well separated in the sky.  However,
we found this not to be the case.  We found that the matrix was diagonal
dominated and mainly noise-dominated in the off-diagonal elements.  We
therefore have attempted no correction for the slice-to-slice correlation.

The covariance matrix described above did contain a small positive constant
offset, possibly indicating a slight correlation among all the slices.  In
other words, deviations of the six slices from the fitted curve were similar to
each other inasmuch as all showed slightly more isolated clusters than voids.
Such an effect could be just a statistical fluctuation or it might be due to
small non-linear effects, as have been found in three dimensions (Matsubara
1996).

There can also be a point-to-point correlation in the genus curve, since the
genus of neighboring $\nu$ values counts some identical structures.  As such,
the number of degrees of freedom which actually resides in our fits is not
necessarily twenty-one, as shown in figure 2---twenty-one samplings of the
genus is a somewhat arbitrary number to choose; we would be surprised if this
exactly matched the number of independent variates in the genus.  In order to
assess the true number of independent variables in the genus, we created 36
simulated groups of 6 slices with Gaussian random phase fluctuations and a
power spectrum as measured by Landy \etal\ (1996) for the LCRS.  When we add
our real dataset, we have 37 independent datasets, 36 of which are guaranteed
to be Gaussian and random-phase.  For each dataset, $m$, we computed the best
fit theoretical curve, and derived its pairwise covariance among the various
values of $\nu$,
\begin{equation}
C_{m,ij} = (\bar{g}_{m,i}-\tilde{g}_{m,i})(\bar{g}_{m,j}-\tilde{g}_{m,j})
\end{equation}
Treating all datasets equivalently, we left out each one in turn, and computed
from the remaining 36 a model covariance matrix as the average of $C_{m,ij}$
over the remaining $m$ values, \ie\ $C_{ij} = \sum{C_{m,ij}}/36$.  This allowed
a direct computation of $\chi^2$ in the 37th dataset (the one that was left
out), which should reveal the true number of degrees of freedom in our genus
curve.
\begin{equation}
\chi^2 = \sum_{i,j=1}^{21}(\bar{g}_i-\tilde{g}_i)C^{-1}_{ij}
(\bar{g}_j-\tilde{g}_j)
\end{equation}
Using all 37 slices as the independent slice, we found that the mean number of
degrees of freedom was broadly peaked at ($17 \pm 8$); this is essentially
equal to 20, the number expected in the simple analysis, and so there is no
real reason to modify our $\tilde{\chi}^2$ treatment above.  However, now that
we have a formal $\chi^2$ for the observed data and for 36 fake datasets, we
may ask where the rank of the real dataset lies among the 37 total datasets.
We obtain for the real dataset a rank of 21, which is at the 57\% interval.  If
we rank all 222 slices separately and sum the ranks of the real data slices, we
find the sum of the ranks to be 641, at the 43\% interval for a sum of six
ranks among 222 possible slots.  In this somewhat more sophisticated treatment,
the data again appear consistent with the Gaussian random-phase model.

\section{Discussion}

We have measured the topology (genus) of the large-scale structure in the
universe as observed in the Las Campanas Redshift Survey.  We have found the
genus of the density fluctuations in the linear regime to be consistent with
Gaussian random phase fluctuations, as predicted by Inflation.  In combination
with the the topology measurements of microwave background as observed by COBE,
there are now two orthogonal projections of the three-dimensional linear-regime
density fluctuation distribution.  Both projections show the fluctuations are
consistent with a Gaussian random phase distribution, within the uncertainty.

In the next several years, the Sloan Digital Sky Survey will, for the first
time, produce a fully three-dimensional map of the the density distribution in
a large volume of the universe, containing roughly $10^6$ redshifts.  This will
allow a definitive understanding of the topology of the initial fluctuations.
Also, MAP and Cobras-Samba satellites will map the two-dimensional distribution
of primordial fluctuations on scales of arcminutes.  Any deviation from
Gaussian random-phase should be easily detectable in these samples.

The prediction of Gaussian random-phase primordial density fluctuations by
inflation theory, has passed the test in yet another observational sample.  The
genus topology of the Las Campanas Redshift Survey provides the strongest test
to date of that prediction.  As the observations improve dramatically in the
next several years, very tight constraints on the fluctuations responsible for
structure formation will emerge.

\section{Acknowledgments}

We thank the LCRS team for its generosity in making their results publicly
available.  J.R. Gott, III, has provided critical guidance and advice
throughout the duration of this work, for which the author is most grateful.
Robert Lupton has been a continual resource for statistical issues to the
author.  WNC thanks the Fannie and John Hertz Foundation for its continued and
gracious support.  This research has been partially supported by NSF Grant
AST-9529120 and by NASA Grant NAG5-2759.

\vfill
\eject

\begin{figure}[p]
\plotfiddle{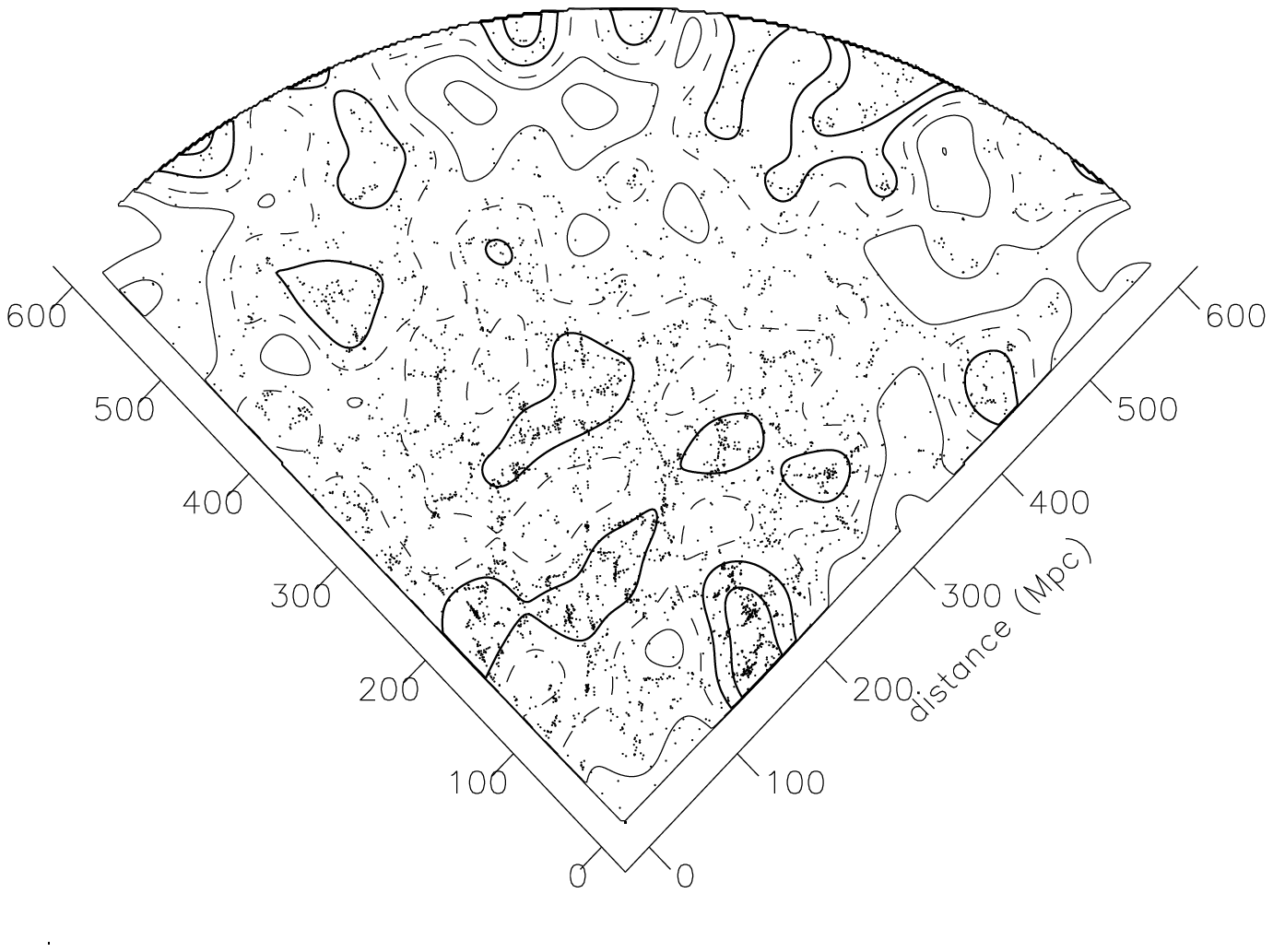}{12cm}{0}{120}{120}{-350}{-460}
\caption{Contours of $\nu = \{-2,-1,0,1,2\}$ in the $\delta = -39\deg$ slice
of the LCRS.  Overplotted are the galaxy locations themselves.}
\end{figure}

\begin{figure}[p]
\plotfiddle{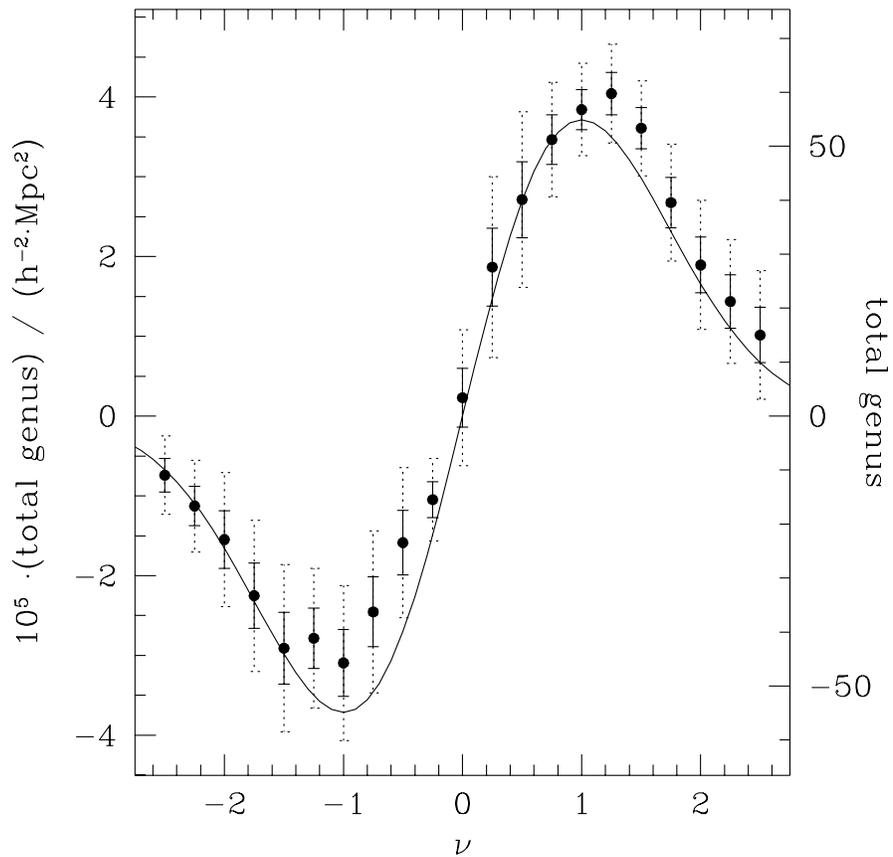}{12cm}{0}{60}{60}{-185}{-90}
\caption{The Genus curve of structure in LCRS.  Smoothing length is $20\hMpc$.
The fit is $G = 6.14\times 10{-5}\nu\exp(-\nu^2/2)$.  Solid errorbars are
68.3\% confidence intervals.  Dotted errorbars are the 95\% confidence
intervals, given by the Student's $t$-distribution.  The plot is also labeled
for the total structures counted in the survey.}
\end{figure}

\end{document}